\documentclass [11pt, one column, a4paper] {article}

\usepackage[dvips]{epsfig}
\usepackage{amsmath}
\usepackage{amssymb}
\usepackage{graphicx}
\usepackage{bm}
\usepackage{epstopdf}
\usepackage[unicode=true,
 bookmarks=true,bookmarksnumbered=false,bookmarksopen=false,
 breaklinks=false,pdfborder={0 0 1},backref=false,colorlinks=false]
 {hyperref}
\usepackage{color}

\title{Magnetoelectric boundary simulated by a Chern-Simons-like model}
\author{H.L. Oliveira$^{1}$\thanks{email: helderluiz10@gmail.com}, L.H.C. Borges$^{2}$\thanks{email: luizhenriqueunifei@yahoo.com.br}, F.E. Barone\thanks{email: frederico.barone@gmail.com}, F.A. Barone$^{1}$\thanks{email: fbarone@unifei.edu.br}\\
{\small $^{1}$IFQ - Universidade Federal de Itajub\'a, Av. BPS 1303, Pinheirinho,}\\
{\small Caixa Postal 50, 37500-903, Itajub\'a, MG, Brazil}\\{\small $^{2}$Universidade Federal do ABC, Centro de Ci\^encias Naturais e Humanas,}\\
{\small Rua Santa Ad\'elia, 166, 09210-170, Santo Andr\'e, SP, Brazil}}

\date {}

\begin {document}

\baselineskip=12pt

\maketitle

\begin{abstract}
In this work we study some physical phenomena that emerge in the vicinity of a magnetoelectric boundary. For simplicity, we restrict to the case of a planar boundary described by a coupling between the gauge field with a planar external Chern-Simons-like potential. The results are obtained exactly. We compute the correction undergone by the photon propagator due to the presence of the Chern-Simons coupling and we investigate the interaction between a stationary point-like charge and the magnetoelectric boundary. In the limit of a perfect mirror, where the coupling constant between the field and the potential diverges, we recover the image method. For a non perfect mirror, we show that we have an attenuated image charge and, in addition, an image magnetic monopole whose field strength does not exhibit the presence of the undesirable and artificial divergences introduced by Dirac strings. We also study the interaction between the plate and a quantum particle with spin. In this case we have a kind of charge-magnetic dipole interaction due to the magnetoelectric properties of the plate.
\end{abstract}

\section{\label{I}Introduction}

In the literature, we can find a wide range of models where external $\delta$-like potentials coupled to fields are used to describe material boundaries. This subject has been of great interest for both experimental and theoretical reasons, once typical experimental setups are composed by material boundaries whose influence on the dinamics of fields cannot be disregarded. In this scenario, we can mention, for instance, the interactions between semi-transparent mirrors and point-like particles \cite{GTFABFEB,FABFEB,LHCBAFFFAB}, investigations which concern the interaction between an atom and a $\delta$-like mirror \cite{PKKM,Russo}, the Casimir energy due to the presence of two $\delta$-like mirrors \cite{BHR,Milton,BorUM,KimballA,BordKD,NRVMH,NRVMMH2,PsRj,FoscoLousada,CapaFoscoLousada,Caval,FABFEB2}, and so on.   

In the work of reference \cite{RussoRede}, in the context of the Casimir Effect, it was proposed a model composed by the Maxwell Lagrangian augmented by a term where the gauge field is coupled to $\delta$-potentials concentrated along parallel planes. The coupling exhibits a Chern-Simons-type form and describes semi-transparent mirrors. In the limit where the coupling constant between the mirrors and the field diverges, the model recovers the presence of perfectly conducting plates. In this specific limit, this model is the same as the one considered previously, in reference \cite{PLA2000}. In reference \cite{Russo} it was also considered the interaction between one of this delta-like plate (with Chern-Simons coupling) and a polarizable atom. In reference \cite{FoscoRemmaggi} this model was considered again to study the Casimir Effect.

However, there are some subjects not yet explored appropriately in the literature for such kind of Chern-Simons-type coupling with a delta-like potential. We can mention, for instance, the meaning of the modifications that the free photon propagator undergoes due to the presence of this single semi-transparent mirror, and the influence of this kind of surface in the dynamics of point-like field sources. In addition, this kind of plate exhibits magnetoelectric properties which were not explored in the literature and could be of some relevance for the analytical description of this metamaterials \cite{PRLmetamateriais,Naturemetamateriais,PRXmetamateriais}.  We hope that the present work could be a contribution to pave the way in this subject.

In this paper we consider some peculiarities of the model poposed in \cite{Russo}. In section \ref{II} we compute the modification undergone by the free photon propagator. We perform our analysis in a similar manner that was employed in Ref. \cite{FABFEB}, where another type of $\delta$-potential was considered. In section \ref{III} we investigate the interaction between a stationary point-like charge and the mirror. We show that the classical image method is found as a particular case of the obtained result. In section \ref{IV} we calculate the electromagnetic field configuration induced by a stationary point-like charge in the presence of the mirror. We show that, in this setup, it emerges the field of a magnetic monopole, without the typical singularity introduced by semi-infinite Dirac strings. It is a new magnetoelectric effect and is an indication that the model could be of some relevance for the description of magnetoelectric metamaterials. In section (\ref{V}) we consider the interaction between the plate and a quantum particle with spin. We show that it emerges a magnetoelectric anisotropic contribution to this energy linear in the spin. Section \ref{conclusoes} is devoted for conclusions and final remarks.

In this paper we work in a $3+1$-dimensional Minkowski space-time  with metric $\eta^{\rho\nu}=(1,-1,-1,-1)$. The Levi-Civita tensor is denoted by $\epsilon^{\rho\nu\alpha\beta}$ with $\epsilon^{0123}=1$.

\section{\label{II} The modified photon propagator}

In this section we consider a model composed by the  Maxwell Lagrangian augmented by a $\delta$-potential concentrated along a plane and with a Chern-Simons-like coupling. The model describes a kind of partially reflective mirror. Without loss of generality, and for convenience, we will consider the mirror perpendicular to the $x^{3}$ axis and placed on the plane $x^{3}=a$. The corresponding model is given by the following Lagrangian density, 
\begin{eqnarray}
\label{LWMirror}
{\cal{L}}&=&-\frac{1}{4}F_{\rho\nu}F^{\rho\nu}-\frac{1}{2\xi}\left(\partial_{\rho}A^{\rho}\right)^{2}\nonumber\\
&
&-\frac{1}{2}\mu v^{\rho}\epsilon_{\rho\nu\alpha\beta}A^{\nu}\partial^{\alpha}A^{\beta}\delta\left(x^{3}-a\right)-J^{\rho}A_{\rho} \ ,
\end{eqnarray}
where $A^{\rho}$ is the photon field, $F^{\rho\nu}=\partial^{\rho}A^{\nu}-\partial^{\nu}A^{\rho}$ is the field strength, $\xi$ is a gauge fixing parameter, $j^{\rho}$ is an external source, 
$v^{\rho}=\eta^{\rho}_{\ 3}=\left(0,0,0,1\right)$ is the normal vector to the surface and $\mu>0$ is a dimensionless coupling constant which accounts for the degree of transparency of the mirror. 

There is a generalization of the Chern-Simons electrodynamics in 3 + 1 dimensions, the so called Carroll-Field-Jackiw model \cite{CFJ}, which exhibits a Lorentz symmetry breaking controlled by a single background 4-vector. The Lagrangian (\ref{LWMirror}) can be seen as a kind of the Carroll-Field-Jackiw Electrodynamics where the Chern-Simons-like term is defined just along a plane perpendicular to the background vector.

The model (\ref{LWMirror}) exhibits a $\delta$-type divergence on the mirror. In order to understand the role of this divergence on the field strength tensor, it is convenient to write explicitly the dynamical equations, as follows
\begin{eqnarray}
\label{motionE1}
\partial_{\rho}F^{\rho\nu}+\mu\delta\left(x^{3}-a\right){\tilde{F}}^{3\nu}=J^{\nu} \ ,
\end{eqnarray}
where we defined ${\tilde{F}}^{\rho\nu}=\frac{1}{2}\epsilon_{\rho\nu\alpha\beta}F^{\alpha\beta}$. For the electric and magnetic fields, we obtain
\begin{eqnarray}
\label{motionE2}
{\bf{\nabla}}\cdot{\bf{E}}&=&J^{0}-\mu\delta\left(x^{3}-a\right)\left({\bf{B}}_{\perp}\cdot{\hat{z}}\right) \ ,\\
\label{motionE3}
{\bf{\nabla}}\times{\bf{B}}&=&{\bf{J}}+ \frac{\partial{\bf{E}}}{\partial t}-\mu\delta\left(x^{3}-a\right)\left({\bf{E}}_{\parallel}\times{\hat{z}}\right) \ ,
\end{eqnarray}
where we defined the vectors perpendicular and parallel to the plate, ${\bf{B}}_{\perp}=\left(0,0,B^{3}\right)$, ${\bf{E}}_{\parallel}=\left(E^{1},E^{2},0\right)$, respectively. 

At this point some comments are in order. The presence of the delta-like potential leads to additional terms in the dynamical equations, which depend on the fields and not on their derivatives. In Eq. (\ref{motionE2}) the delta-like term is proportional to the component of the magnetic field perpendicular to the mirror and can be interpreted as an extra source for the divergence of the electric field. In the same way, in Eq. (\ref{motionE3}) the delta-like term is an additional source for the curl of the magnetic field and depends on the components of the electric field parallel to the surface. For the potential considered in Ref. \cite{FABFEB} the delta-like contributions for the dynamical field equations exhibit different behaviors in comparison with (\ref{motionE1}), (\ref{motionE2}) and (\ref{motionE3}) and depend on the derivatives of the fields. 

Neglecting surface terms, one can write
\begin{eqnarray}
\label{model1}
{\cal{L}}=\frac{1}{2}A^{\rho}{\cal{O}}_{\rho\nu}A^{\nu}-J^{\nu}A_{\nu} \ ,
\end{eqnarray}
where ${\cal{O}}_{\rho\nu}$ is a differential operator. 

For our purpose, it is convenient to split ${\cal{O}}_{\rho\nu}$ into two parts, one corresponding to the usual photon operator and the other one corresponding to the $\delta$-like term, as follows
\begin{eqnarray}
\label{model2}
{\cal{O}}_{\rho\nu}={\cal{O}}^{(0)}_ {\rho\nu}+ \Delta{\cal{O}}_{\rho\nu} \ ,
\end{eqnarray}
with
\begin{eqnarray}
\label{operator1}
{\cal{O}}^{(0)}_ {\rho\nu}&=&\eta_{\rho\nu}\Box \ ,\\
\label{operator2}
\Delta{\cal{O}}_{\rho\nu}&=&-{\mu}\delta\left(x^{3}-a\right)\epsilon_{3\rho\alpha\nu}\partial_{\parallel}^{\alpha} \ ,
\end{eqnarray}
where we defined the operator $\Box=\partial^{\alpha}\partial_{\alpha}$ and used a gauge where $\xi=1$. We notice that the derivative in (\ref{operator2}) is defined only in the Minkowski coordinates parallel to the mirror, namely, $\partial_{\parallel}^{\alpha}=\left(\partial^{0},\partial^{1},\partial^{2}\right)$, because of the fixed index in the Levi-Civita tensor.

The free photon propagator satisfies the differential equation ${\cal{O}}^{(0)\rho\nu}(x)G^{(0)}_{\nu\beta}\left(x,y\right)=\eta^{\rho}_{\ \beta}\delta^{4}\left(x-y\right)$ and is given in the Feynman gauge, $(\xi=1)$, by 
\begin{eqnarray}
\label{prop0}
G^{(0)}_{\rho\nu}\left(x,y\right)=-\eta_{\rho\nu}\int\frac{d^{4}p}{\left(2\pi\right)^{4}}\frac{e^{-ip\cdot\left(x-y\right)}}{p^{2}}\ .
\end{eqnarray}

The propagator corresponding to the model (\ref{model1}) must satisfy the equation ${\cal{O}}^{\rho\nu}(x)G_{\nu\lambda}\left(x,y\right)=\eta^{\rho}_{\ \lambda}\delta^{4}\left(x-y\right)$ and can be obtained recursively \cite{FABFEB,GTFABFEB,FABFEB2,LHCBAFFFAB,Caval}. It is simple to check that 
\begin{eqnarray}
\label{prop1}
&&G_{\rho\nu}\left(x,y\right)=G^{(0)}_{\rho\nu}\left(x,y\right)\nonumber\\
&
&-\int d^{4}z \ G_{\rho\gamma}\left(x,z\right)\Delta{\cal{O}}^{\gamma\sigma}\left(z\right)G^{(0)}_{\sigma\nu}\left(z,y\right) \ ,
\end{eqnarray}

For convenience let us write $G_{\rho\nu}\left(x,y\right)$ and $G^{(0)}_{\rho\nu}\left(x,y\right)$ as Fourier integrals in the parallel coordinates, as follows
\begin{eqnarray}
\label{prop2}
G_{\rho\nu}\left(x,y\right)=\int\frac{d^{3}p_{\parallel}}{\left(2\pi\right)^{3}} \ {\cal{G}}_{\rho\nu}\left(x^{3},y^{3};p_{\parallel}\right)e^{-ip_{\parallel}\cdot\left(x_{\parallel}-y_{\parallel}\right)} , \\
\label{prop3}
G^{(0)}_{\rho\nu}\left(x,y\right)=\int\frac{d^{3}p_{\parallel}}{\left(2\pi\right)^{3}} \ {\cal{G}}^{(0)}_{\rho\nu}\left(x^{3},y^{3};p_{\parallel}\right)e^{-ip_{\parallel}\cdot\left(x_{\parallel}-y_{\parallel}\right)} ,
\end{eqnarray}
where we defined $x_{\parallel}^{\rho}=\left(x^{0},x^{1},x^{2}\right)$ and $p_{\parallel}^{\rho}=\left(p^{0},p^{1},p^{2}\right)$ as being the coordinates and momenta parallel to the mirror, respectively. The functions ${\cal{G}}_{\rho\nu}\left(x^{3},y^{3};p_{\parallel}\right)$ and  ${\cal{G}}^{(0)}_{\rho\nu}\left(x^{3},y^{3};p_{\parallel}\right)$ are usually known as reduced Green's functions \cite{FABFEB,GTFABFEB,FABFEB2,LHCBAFFFAB,Caval}.

Using the fact that \cite{ISGrad}
\begin{eqnarray}
\label{intperp}
\int\frac{dp^{3}}{\left(2\pi\right)}\frac{e^{ip^{3}\left(x^{3}-y^{3}\right)}}{p^{2}}=-\frac{e^{-\sigma\mid x^{3}-y^{3}\mid}}{2\sigma} \ ,
\end{eqnarray}
with $p^{3}$ standing for the momentum perpendicular to the mirror and $\sigma=\sqrt{-p_{\parallel}^{2}}$, from  Eq. (\ref{prop0}) we have for the reduced photon propagator 
\begin{eqnarray}
\label{prop4}
{\cal{G}}^{(0)}_{\rho\nu}\left(x^{3},y^{3};p_{\parallel}\right)&=& \eta_{\rho\nu}\frac{e^{-\sigma\mid x^{3}-y^{3}\mid}}{2\sigma} \ .
\end{eqnarray}

Substituting (\ref{operator2}) into (\ref{prop1}), using Eqs. (\ref{prop2}), (\ref{prop3}) and (\ref{prop4}),  and performing some simple integrals, we obtain
\begin{eqnarray}
\label{prop5}
&{\cal{G}}_{\rho\nu}&\left(x^{3},y^{3};p_{\parallel}\right)=\eta_{\rho\nu}\frac{e^{-\sigma\mid x^{3}-y^{3}\mid}}{2\sigma}\nonumber\\
&
&-i\mu \ \epsilon_{3\ \  \nu}^{\ \gamma\alpha} \ {\cal{G}}_{\rho\gamma}\left(x^{3},a;p_{\parallel}\right)p_{\parallel\alpha}\frac{e^{-\sigma\mid y^{3}-a\mid}}{2\sigma} \ .
\end{eqnarray}

From the above equation, the reduced Green's function ${\cal{G}}_{\rho\nu}\left(x^{3},y^{3};p_{\parallel}\right)$ must be obtained recursively. Evaluating Eq. (\ref{prop5}) for $y^{3}=a$ and performing some simple manipulations, we arrive at
\begin{eqnarray}
\label{prop6}
{\cal{G}}_{\rho\gamma}\left(x^{3},a;p_{\parallel}\right)\left(\eta_{\ \nu}^{\gamma}+\frac{i\mu}{2\sigma} \ \epsilon_{3\ \  \nu}^{\ \gamma\alpha} \ p_{\parallel\alpha}\right)=\eta_{\rho\nu}\frac{e^{-\sigma\mid x^{3}-a\mid}}{2\sigma} \ .\cr
\ 
\end{eqnarray}

Now we multiply both sides of (\ref{prop6}) by the inverse of the operator $\eta_{\ \nu}^{\gamma}+\frac{i\mu}{2\sigma} \ \epsilon_{3\ \  \nu}^{\ \gamma\alpha} \ p_{\parallel\alpha}$, what leads to
\begin{eqnarray}
\label{prop7}
&&{\cal{G}}_{\rho\gamma}\left(x^{3},a;p_{\parallel}\right)=
\frac{e^{-\sigma\mid x^{3}-a\mid}}{2\sigma}\Biggl[\eta_{\rho\gamma}\nonumber\\
&
&-\frac{\mu^{2}}{\mu^{2}+4}\left(\eta_{\parallel\rho\gamma}-\frac{p_{\parallel\rho}p_{\parallel\gamma}}{p_{\parallel}^{2}}\right)-\frac{2i\mu}{\sigma\left(\mu^{2}+4\right)}\epsilon_{3\rho\alpha\gamma}  p_{\parallel}^{\alpha}\Biggr] \ , \nonumber\\
\end{eqnarray}
where we defined  $\eta_{\parallel}^{\alpha\beta}=\eta^{\alpha\beta}-\eta^{\alpha 3}\eta^{\beta 3}$.

Substituting Eq. (\ref{prop7}) in (\ref{prop5}) and using the Eq. (\ref{prop2}), the modified photon propagator due to the presence of the semi-transparent mirror reads
\begin{eqnarray}
\label{prop10}
&{{G}}_{\rho\nu}\left(x,y\right)&=\int\frac{d^{3}p_{\parallel}}{\left(2\pi\right)^{3}} \ e^{-ip_{\parallel}\cdot\left(x_{\parallel}-y_{\parallel}\right)}\Biggl\{\eta_{\rho\nu}\frac{e^{-\sigma\mid x^{3}-y^{3}\mid}}{2\sigma}\nonumber\\
&
&-\frac{e^{-\sigma\left(\mid x^{3}-a\mid +\mid y^{3}-a\mid\right)}}{4+\mu^{2}}\Biggl[\frac{\mu^{2}}{2\sigma}\left(\eta_{\parallel\rho\nu}-\frac{p_{\parallel\rho}p_{\parallel\nu}}{p_{\parallel}^{2}}\right)\nonumber\\
&
&+\frac{i\mu}{p_{\parallel}^{2}}\epsilon_{3\alpha\rho\nu}p_{\parallel}^{\alpha}\Biggr]\Biggr\} \ .
\end{eqnarray}

The propagator (\ref{prop10}) is continuous and well defined all over the space, except at coincident points, where $x^{\mu}=y^{\mu}$. The first term on the right-hand side is just the free photon propagator. The second and third terms are corrections which account for the presence of the semi-transparent mirror. As an important check, we point out that by taking the limit $\mu\rightarrow\infty$ in Eq. (\ref{prop10}) we recover the well-known photon propagator due to the presence of a single perfect mirror \cite{Bordag}.

\section{\label{III} Charge-mirror interaction}

In the present section we consider the interaction energy between a stationary point-like charge and the semi-transparent mirror studied in the previous section. Since we have a quadratic Lagrangian in the field variables, as discussed in references \cite{
FABGFHPrd,BJPBaroneHidalgo,KalbRamond,FABAAN2,LHCFABHel,FABAAN1,LHCFABAFFEpjc,LHCBFABplate,Medeiros,LHCFABBjp,LHCBFABplate2,Oliveira}, the contribution due to the external sources to the ground state energy of the system is given by 
\begin{eqnarray}
\label{energy}
E=\frac{1}{2T}\int d^{4}x \ d^{4}y \ J^{\rho}\left(x\right)
{{G}}_{\rho\nu}\left(x,y\right)J^{\nu}\left(y\right) \ ,
\end{eqnarray}
where $T$ is the time variable and it is implicit the limit $T\to\infty$.

With no loss of generality, we choose a charge placed at position ${\bf b}=\left(0,0,b\right)$. The corresponding external source reads
\begin{eqnarray}
\label{source1}
J^{\rho}\left(x\right)=q\eta^{\rho 0}\delta^{3}\left({\bf x}-{\bf b}\right) \ ,
\end{eqnarray}
where the parameter $q$ is the electric charge intensity.

We notice that the first term on the right-hand side of expression (\ref{prop10}) comes from the free photon propagator (without the presence of the mirror) and, therefore, it does not  contribute to the interaction energy between the charge and the mirror.  Indeed, this contribution does not depend on the distance between the charge and the mirror, it is present even in the absence of the mirror and provides the charge self energy. Thus, only the last two terms of propagator contribute to the interaction energy.

Substituting Eqs. (\ref{source1}) and (\ref{prop10}) in (\ref{energy}), discarding self interacting terms and performing the integrals in $d^{3}{\bf x}$, $d^{3}{\bf y}$, $dx^{0}$, $dp^{0}$, $dy^{0}$, we obtain 
\begin{eqnarray}
\label{energyC}
E_{MC}\left(R,\mu\right)=-\frac{q^{2}}{16\pi^{2}}\left(\frac{\mu^{2}}{4+\mu^{2}}\right)\int d^{2}{\bf{p}}_{\parallel}\frac{e^{-2R\sqrt{{\bf{p}}_{\parallel}^{2}}}}{\sqrt{{\bf{p}}_{\parallel}^{2}}} \ ,
\end{eqnarray}
where $R=\mid a-b\mid$ is the distance between the mirror and the charge.
The sub-index $MC$ means that we have the interaction energy between the mirror and
the charge.

Using polar coordinates and performing some simple integrations, the interaction energy reads
\begin{eqnarray}
\label{energyC2}
E_{MC}\left(R,\mu\right)=-\frac{q^{2}}{16\pi R}\left(\frac{\mu^{2}}{4+\mu^{2}}\right) \ .
\end{eqnarray}

\begin{figure}[!h]
\centering \includegraphics[scale=0.5]{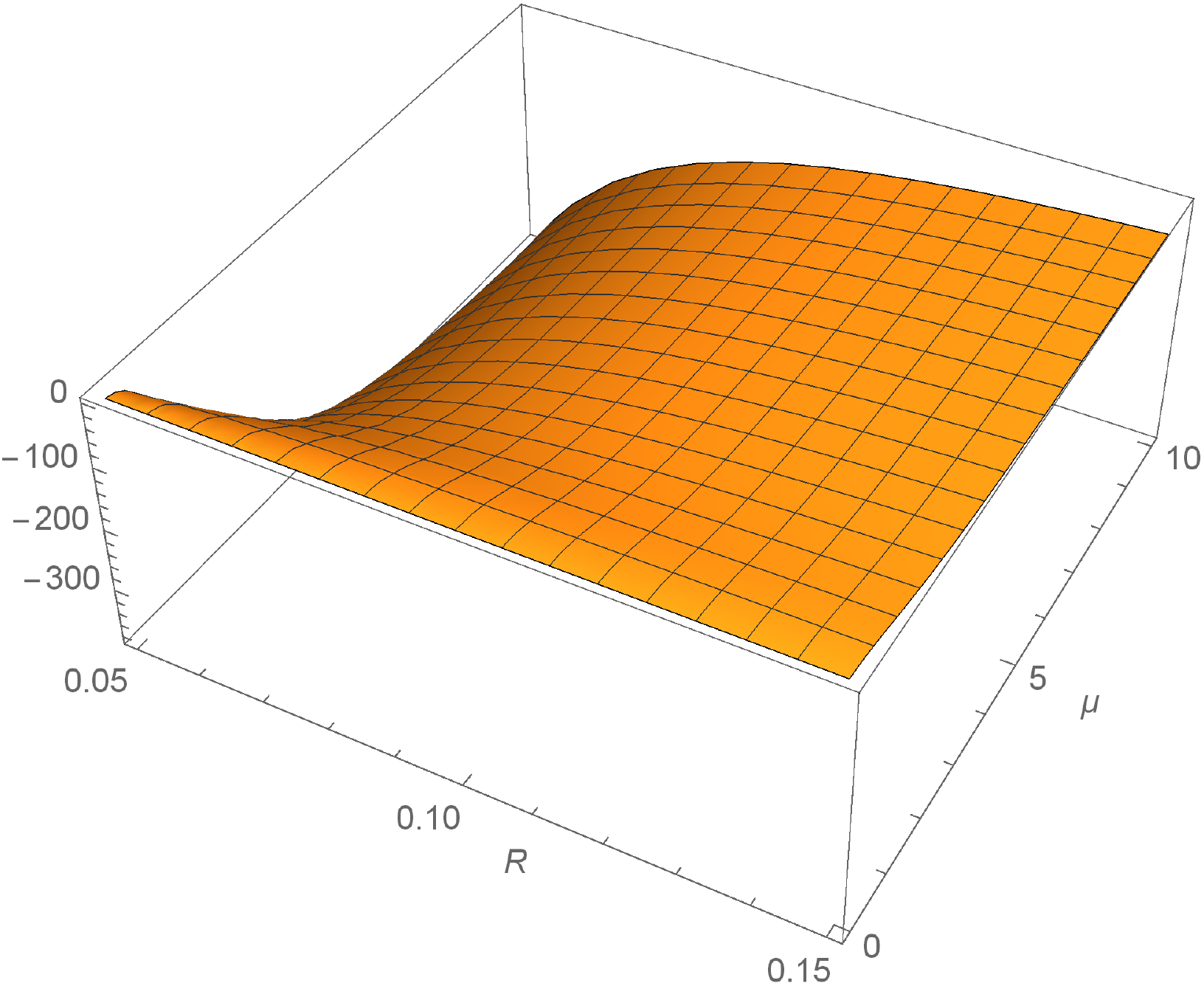} \caption{ Force (\ref{forcem}) multiplied by $\frac{16\pi}{q^{2}}$.}
\label{grafico1} 
\end{figure}

Eq. (\ref{energyC2}) is the exact result for the interaction energy between a point-like charge and the two-dimensional semi-transparent mirror, described by the model (\ref{LWMirror}). It is interesting to realize that the energy (\ref{energyC2}) still exhibits a Coulomb-like behavior. The role played by the coupling constant $\mu$ is just to attenuate the image charge ($q_{im}$) by an uniform multiplicative factor, namely, $q_{im}=q\frac{\mu^{2}}{4+\mu^{2}}$. As expected,  this energy vanishes in the limit $\mu\rightarrow 0$, where we have no mirror present. In the limit $\mu\rightarrow\infty$, we have the gauge field subjected to the boundary conditions imposed by a perfect mirror and
\begin{eqnarray}
\label{energyminfty}
\lim_{\mu\rightarrow\infty}E_{MC}\left(R,\mu\right)=-\frac{q^{2}}{16\pi R} \ ,
\end{eqnarray}
what recovers the image method. Thus, the model (\ref{LWMirror}) describes the interaction obtained via image method in the limit of a perfect conductor.

The force between the point-like charge and the mirror is given by
\begin{eqnarray}
\label{forcem}
F_{MC}\left(R,\mu\right)=-\frac{\partial}{\partial R}E_{MC}\left(R,\mu\right)=-\frac{q^{2}}{16\pi R^{2}}\frac{\mu^{2}}{4+\mu^{2}}\ ,\cr
\ 
\end{eqnarray}
which is always negative and, therefore, has an attractive behavior. In the figure (\ref{grafico1}) we have a plot for the force (\ref{forcem}) multiplied by $\frac{16\pi}{q^{2}}$ as a function of $R$ and $\mu$. 

Here, one comment is in order. The parameter $\mu$ is dimensionless. In this case, by a dimensional analysis, we could infer the $q^{2}/R$ behavior for the energy (\ref{energyC2}) previously, just before we perform the calculations. The numerical factor  $-1/(16\pi)$, including the minus sign, and the dependence on $\mu$ cannot be obtained before the calculations are performed.

\begin{figure}[!h]
\centering \includegraphics[scale=0.38]{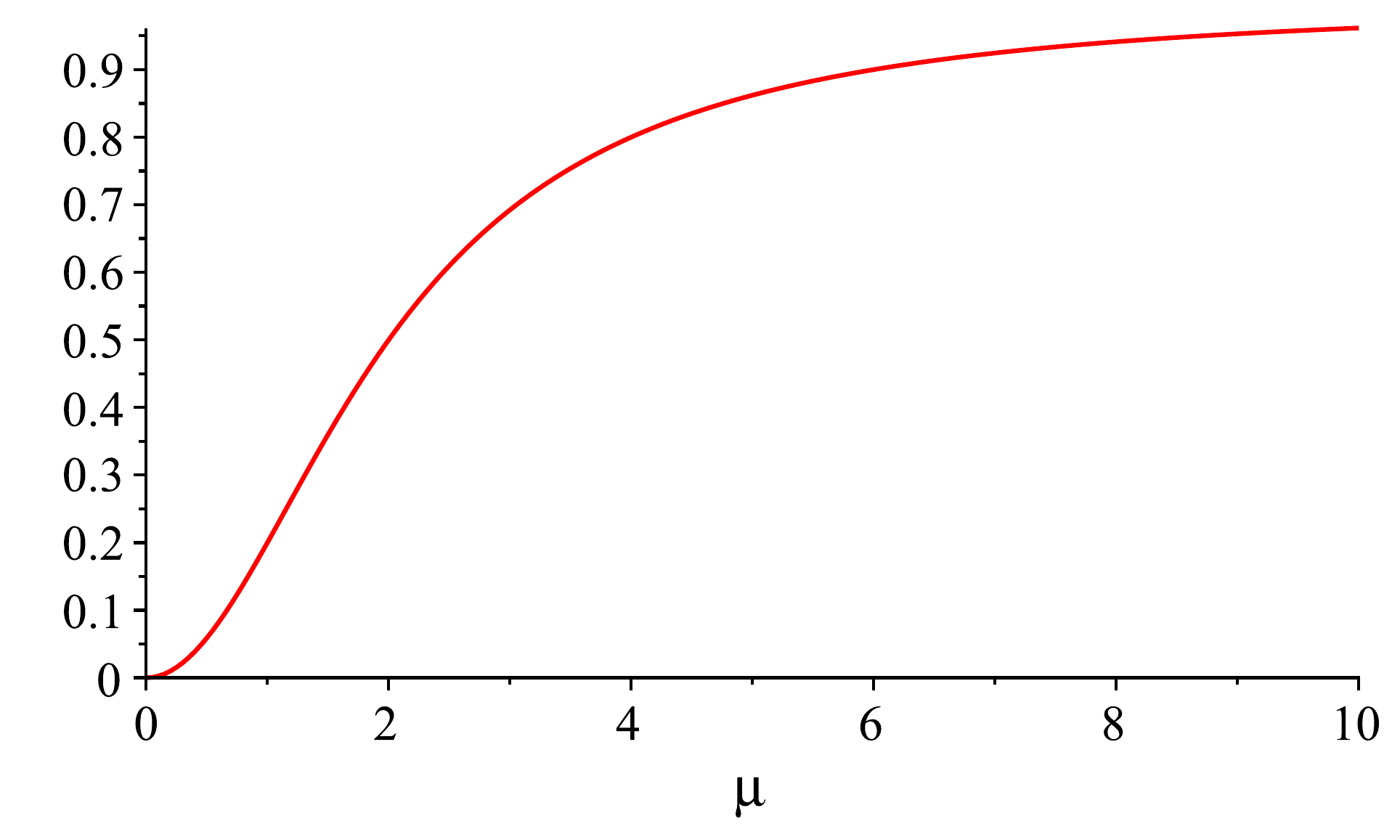} \caption{Dimensionless numerical factor $\mu^{2}/(4+\mu^{2})$ (vertical axis) as a function of $\mu$ (horizontal axis).}
\label{fatormu} 
\end{figure}

The dimensionless numerical factor $\mu^{2}/(4+\mu^{2})$ lies in the range $0\leq\mu^{2}/(4+\mu^{2})\leq1$, as we can see in figure (\ref{fatormu}), so the force (\ref{forcem}) as well as the energy (\ref{energyC2}), attain their minimum values for a perfectlly conducting plate.

\section{\label{IV} Electromagnetic field}

In this section we calculate the electromagnetic field configuration produced by a stationary point-like charge in the vicinity of the semi-transparent magnetoelectric mirror. We start by considering the particular solution for the electromagnetic four-potential in the presence of an external source 
\begin{eqnarray}
\label{EMF33}
A_{\rho}\left(x\right)=\int d^{4}y \  G_{\rho\nu}\left(x,y\right)J^{\nu}\left(y\right) \ , 
\end{eqnarray}
where $G^{\rho\nu}\left(x,y\right)$ is the propagator (\ref{prop10}) and $J_{\nu}\left(y\right)$ is given by (\ref{source1}).

From now on, with no loss of generality, we shall take a coordinate system where $b^{3}>0$ and $a=0$. It means that the mirror is placed along the plane $x^{3}=0$ and the particle is placed at a given point on the positive $x^{3}$ axis.

Each term on the right hand side of (\ref{prop10}) leads to a different contribution to the 4-vector potential in (\ref{EMF33}). The first term is the one obtained from the free photon propagator (without the presence of the mirror),
\begin{eqnarray}
\int d^{4}y\frac{d^{3}p_{\parallel}}{\left(2\pi\right)^{3}} \ e^{-ip_{\parallel}\cdot(x_{\parallel}-y_{\parallel})}\eta_{\rho\nu}\frac{e^{-\sigma\mid x^{3}-y^{3}\mid}}{2\sigma}q\eta^{\nu0}\delta^{3}({\bf y}-{\bf b})\cr\cr
=q\eta_{\rho}^{\ 0}\int\frac{d^{2}p_{\parallel}}{\left(2\pi\right)^{2}} \ e^{i{\bf p}_{\parallel}\cdot{\bf x}_{\parallel}}\frac{e^{-\sqrt{{\bf p}_{\parallel}^{2}}\mid x^{3}-y^{3}\mid}}{2\sqrt{{\bf p}_{\parallel}^{2}}}\ .\ \ \ \ 
 \end{eqnarray}

Defining as $\varphi$ the angle between the vectors ${\bf p}_{\parallel}$ and ${\bf x}_{\parallel}$ in the ${\bf p}_{\parallel}$ space and $s=\sqrt{{\bf p}_{\parallel}^{2}}$, we can write the above integral in cylindrical coordinates as follows
\begin{eqnarray}
\label{contr1}
\int\!\!\! &d^{4}y&\!\!\!  \frac{d^{3}p_{\parallel}}{\left(2\pi\right)^{3}} \ e^{-ip_{\parallel}\cdot(x_{\parallel}-y_{\parallel})}\eta_{\rho\nu}\frac{e^{-\sigma\mid x^{3}-y^{3}\mid}}{2\sigma}q\eta^{\nu0}\delta^{3}({\bf y}-{\bf b})\cr\cr
&=&\frac{q\eta_{\rho}^{\ 0}}{(2\pi)^2}\int_{0}^{\infty}ds\int_{0}^{2\pi}d\varphi\
\frac{e^{-s|x^{3}-b^{3}|}}{2}
e^{is|{\bf x}_{\parallel}|\cos\varphi}\cr\cr
&=&\frac{q}{4\pi}\frac{\eta_{\rho}^{\ 0}}{\sqrt{|{\bf x}_{\parallel}|^{2}+|x^{3}-b^{3}|^{2}}}
=\frac{q}{4\pi}\frac{\eta_{\rho}^{\ 0}}{|{\bf x}-{\bf b}|}\ ,
\end{eqnarray}
what is the well known potential of a stationary electric charge $q$ placed at ${\bf b}$.

The second term of the propagator (\ref{prop10}) gives the following contribution to the 4-potential (\ref{EMF33})
\begin{eqnarray}
\label{contr2}
\int d^{4}y\frac{d^{3}p_{\parallel}}{\left(2\pi\right)^{3}} -\frac{\mu^{2}}{4+\mu^{2}} \left(\eta_{\parallel\rho\nu}-\frac{p_{\parallel\rho}p_{\parallel\nu}}{p_{\parallel}^{2}}\right)\cr\cr
e^{-ip_{\parallel}\cdot(x_{\parallel}-y_{\parallel})}\frac{e^{-\sigma\left(\mid x^{3}-a\mid +\mid y^{3}-a\mid\right)}}{2\sigma}q\eta^{\nu0}\delta^{3}({\bf y}-{\bf b})\cr\cr
=-\frac{q\eta_{\rho}^{\ 0}}{(2\pi)^2}\frac{\mu^{2}}{4+\mu^{2}} \int_{0}^{\infty}\!\!\! ds\int_{0}^{2\pi}\!\!\! d\varphi\ 
\frac{e^{-s(|x^{3}|+|b^{3}|)}}{2}e^{is|{\bf x}_{\parallel}|\cos\varphi}\cr\cr
=-\frac{q\eta_{\rho}^{\ 0}}{4\pi}\frac{\mu^{2}}{4+\mu^{2}}
\frac{1}{|{\bf x}+sgn(x^{3}){\bf b}|}\ .\ \ \ \ \ \ 
\end{eqnarray}
where we defined the sign function $sgn(x>0)=1$; $sgn(x<0)=-1$ and $sgn(0)=0$.

The contribution of the third term in (\ref{prop10}) to the field (\ref{EMF33}) must be considered cautiously, as follows

\begin{eqnarray}
\label{contr3}
-\frac{i\mu}{4+\mu^{2}}
\int d^{4}y\frac{d^{3}p_{\parallel}}{\left(2\pi\right)^{3}}
e^{-ip_{\parallel}\cdot\left(x_{\parallel}-y_{\parallel}\right)}
\frac{e^{-\sigma\left(\mid x^{3}\mid +\mid y^{3}\mid\right)}}{2\sigma}\cr\cr
\times\epsilon_{3\alpha\rho\nu}\frac{p_{\parallel}^{\alpha}}{p_{\parallel}^{2}}q\delta^{3}({\bf y}-{\bf b})\cr\cr
=\frac{q\ \mu}{4+\mu^{2}}
\epsilon_{3j\rho0}\left(\frac{\partial|{\bf x}_{\parallel}|}{\partial x^{j}}\right)\int_{0}^{\infty}\frac{ds}{2\pi}
\frac{e^{-s\left(\mid x^{3}\mid +\mid b^{3}\mid\right)}}{s}\cr\cr
\frac{\partial}{\partial|{\bf x}_{\parallel}|}
\int_{0}^{2\pi}\frac{d\varphi}{2\pi}e^{is|{\bf x}_{\parallel}|\cos\varphi}\cr\cr
=\frac{q}{2\pi}\frac{\mu\ \epsilon_{0\rho j3}}{4+\mu^{2}}\frac{x_{\parallel}^{j}}{|{\bf x}_{\parallel}|^{2}}
\Bigg[1-\frac{|x^{3}|+b^{3}}{\sqrt{{\bf x}_{\parallel}^{2}+[x^{3}+sgn(x^{3}) b^{3}]^{2}}}\Bigg].
\end{eqnarray}

Collecting the terms (\ref{contr1}), (\ref{contr2}) and (\ref{contr3}), using cylindrical coordinates: $\rho=|{\bf x}_{\parallel}|$, $\phi=arctan(\rho/x^{3})$ and performing some simple algebra, we have the potencial for the mirror-charge system
\begin{eqnarray}
\label{potenciaisparacarga}
A^{0}({\bf x})&=&\frac{q}{4\pi}\Bigg[\frac{1}{|{\bf x}-{\bf b}|}
-\frac{\mu^{2}}{4+\mu^{2}}\frac{1}{|{\bf x}+sgn(x^{3}) {\bf b}|}\Bigg]\cr\cr
{\bf A}({\bf x})&=&\frac{q}{4\pi}\frac{2\mu}{4+\mu^{2}}
\Bigg[1-sgn(x^{3})\frac{x^{3}+sgn(x^{3})b^{3}}{|{\bf x}+sgn(x^{3}) {\bf b}|}\Bigg]\frac{\hat\phi}{\rho}\ .\cr
&\ &\ 
\end{eqnarray}
from which we obtain the fields
\begin{eqnarray}
\label{EMF7}
{\bf E}\left({\bf{x}}\right)&=&\frac{q}{4\pi}\frac{{\bf x}-{\bf b}}{|{\bf x}-{\bf b}|^{3}}
-\frac{q}{4\pi}\left(\frac{\mu^{2}}{4+\mu^{2}}\right)
\frac{{\bf x}+sgn(x^{3}){\bf b}}{|{\bf x}+sgn(x^{3}){\bf b}|^{3}}\cr\cr\cr
{\bf B}\left({\bf{x}}\right)&=&\frac{q}{4\pi}\left(\frac{2\mu}{4+\mu^{2}}\right)sgn(x^{3})
\frac{{\bf x}+sgn(x^{3}){\bf b}}{|{\bf x}+sgn(x^{3}){\bf b}|^{3}}\ .
\end{eqnarray}

Here, some comments are in order. In the side of the mirror where the charge $q$ is placed, $x^{3}>0$ and the field $A^{0}({\bf x})$ in (\ref{potenciaisparacarga}) is the potential produced by the charge itself, placed at ${\bf b}$, and an image charge with intensity $-q\mu^{2}/(4+\mu^{2})$ and placed at position $-{\bf b}$. In the opposite side, $x^{3}<0$ and the field  $A^{0}({\bf x})$ reduces to the one produced by a charge placed at ${\bf b}$ and with intensity $q[1-\mu^{2}/(4+\mu^{2})]$. The field $A^{0}({\bf x})$ in (\ref{potenciaisparacarga}) gives the electric field in (\ref{EMF7}). 

In the side of the mirror where the charge is placed, the vector potential ${\bf A}({\bf x})$ in (\ref{potenciaisparacarga}) is the one related to a Dirac monopole with intensity $2q\mu/(4+\mu^{2})$, placed at $-{\bf b}$ and whose corresponding semi-infinite Dirac string is lying on the axis $-\infty<x^{3}<-b^{3}$, which is in the opposite side of the mirror. So, we do not have the typical divergence commonly associated with the string in the description of magnetic monopoles. In the opposite side of the mirror where the charge is placed, the vector potential in (\ref{potenciaisparacarga}) is the same as the one produced by a Dirac monopole placed at ${\bf b}$, with intensity $-2q\mu/(4+\mu^{2})$ and related to a semi-infinite Dirac string lying along the axis $b^{3}<x^{3}<\infty$. Once again, there is no divergence for the potential imposed by the string. The field ${\bf A}({\bf x})$ in (\ref{potenciaisparacarga}) gives the monopole magnetic field in (\ref{EMF7}) which does not exhibit any divergence usually imposed by Dirac strings (a delta-like magnetic monopole) \cite{Felsagger,Camilo,Helayel}. 

It is important to reinforce that the flux of the magnetic field through a closed surface is always zero, for surfaces placed in any side of the mirror. The magnetic flux is also zero through surfaces which encloses a portion of the plate as well. It is due to the fact that we do not have a real magnetic monopole in any situation. All the magnetic monopoles are virtual ones, lying on the opposite side of the mirror were the field is taken. 

In the limit $\mu\to\infty$, the field $A^{0}({\bf x})$ in (\ref{potenciaisparacarga}) and the electric field in (\ref{EMF7}) reduce to the ones obtained for a perfect conductor, and the vector potential ${\bf A}({\bf x})$ in (\ref{potenciaisparacarga}) vanishes, as well as the magnetic field in (\ref{EMF7}). So we recover the same results obtained from a perfect conductor in the limit $\mu\to\infty$.

When $\mu=0$, we recover the case with just a single point-like charge, where we have only a coulombian electric field,  as it should be.

It is very interesting to notice that the modulus of the magnetic field connected to the magnetic image monopole, given by the second expression in Eq. (\ref{EMF7}), can be greater or lower than the modulus of the electric field of the image charge, given by the second term on the right hand side of the first Eq. (\ref{EMF7}) . In fact, from Eqs. (\ref{EMF7}), we have
\begin{equation}
|{\bf E}_{image}|=\frac{\mu}{2}|{\bf B}|\ .
\end{equation}
So when $\mu=2$, the image charge and image monopole produce fields with the same intensities. When $\mu>2$ the image electric field dominates over the image magnetic field. For $\mu<2$, we have the opposite situation.  

\begin{figure}[!h]
\centering \includegraphics[scale=0.4]{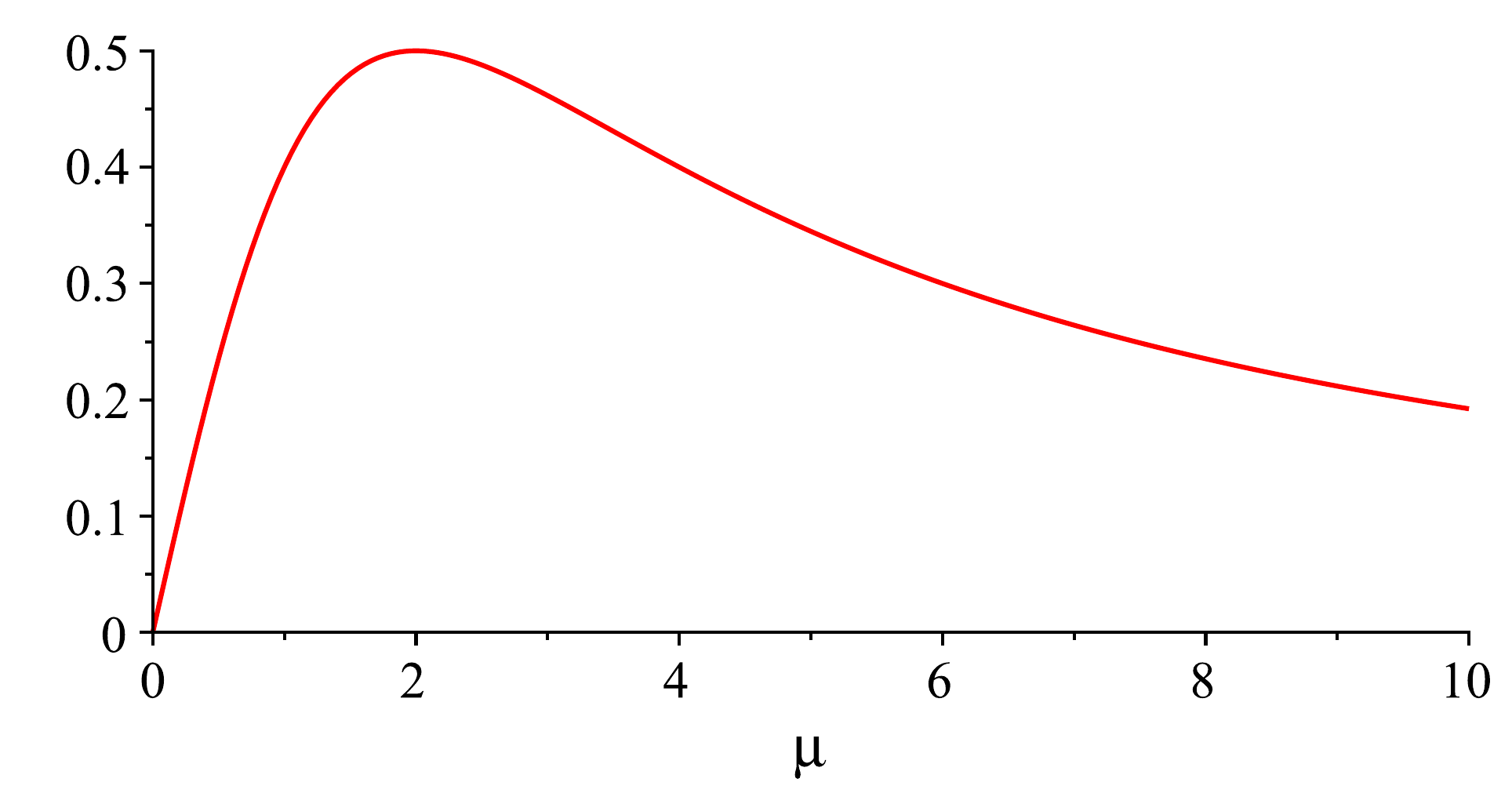} \caption{Dimensionless numerical factor $2\mu/(4+\mu^{2})$ (vertical axis) as a function of $\mu$ (horizontal axis).}
\label{fatormu2} 
\end{figure}

In figure (\ref{fatormu}) we have the graphic for the attenuation factor $\mu^{2}/(4+\mu^{2})$ for the image electric charge. It is a monotonic function of the coupling constant $\mu$. In figure (\ref{fatormu2}) we have the attenuation factor $2\mu/(4+\mu^{2})$ for the image magnetic monopole. It is always positive, equal to zero for $\mu=0$ and $\mu=\infty$, and attains its maximum value, $1/2$, for $\mu=2$.

\section{Plate-spin interaction}
\label{V}

In this section we consider the interaction energy between a charged quantum particle with spin $1/2$ and the plate. This kind of interaction is known for standard situations, but the novelty in this case relies on the fact that we have a magnetoelectric plate. So, for a particle with spin, we shall have a cross interaction between the charge and spin of the particle due to its mirrored image in the magnetoelectric plate.

We shall take a simplified model, where the particle is described by the well known Pauli spin theory. We shall also take the particle as fixed, so its quantum state is just described by spin variables. The quantum states are denoted by $|+\rangle$ and $|-\rangle$, standing for spin up and down, respectively, with the ${\hat z}$ taken as the quantization axis. 

Let us start by considering the interaction energy between the magnetoelectric plate and a classical charged particle with magnetic moment ${\bf m}$. In this case, the field source which accounts for the presence of such a particle is
\begin{equation}
\label{defJCMD}
J_{C,MD}^{\mu}({\bf x})=q\eta^{\mu 0}\delta^{3}({\bf{x}}-{\bf{b}})+\epsilon^{0\mu j k}m_{j}\left[\partial_{k}\delta^{3}({\bf{x}}-{\bf{b}})\right] \ .
\end{equation}
The first term on the right hand side of (\ref{defJCMD}) is, obviously, the source for a stationary point-like charge placed at ${\bf x}={\bf b}$. The second term is the source for a stationary magnetic dipole ${\bf m}$ placed at ${\bf x}={\bf b}$, as discussed in the appendix.
From now on, with no loss of generality, we shall take ${\bf b}=(0,0,b)$.

Substituting Eq. (\ref{defJCMD}) in (\ref{energy}) and performing some straightforward calculations, similar to the ones of the previous sections, we are taken to the interaction energy 
\begin{eqnarray}
\label{ETOTAL}
E&=&\left[-\frac{q^{2}}{16\pi\mid b\mid}+\frac{\left[{\bf{m}}^{2}+\left({\bf{m}}\cdot{\hat{z}}\right)^{2}\right]}{64\pi\mid b\mid^{3}}\right]\left(\frac{\mu^{2}}{4+\mu^{2}}\right)\cr
&\ &-\frac{q\left({\bf{m}}\cdot{\hat{z}}\right)}{8\pi\mid  b\mid^{2}}\left(\frac{\mu}{4+\mu^{2}}\right) \ .
\end{eqnarray}

The first two terms on the right hand side of Eq. (\ref{ETOTAL}) are the direct interactions, that is, the charge $\times$ image-charge and the magnetic dipole $\times$ image-magnetic dipole ones, respectively. In comparison with the case of a perfect mirror, we have an atenuation factor $\mu^{2}/(4+\mu^{2})$ for these first two terms. The last term in (\ref{ETOTAL}) is a cross interaction due to the simultaneous presence of, both, the charge and the magnetic moment. Notice that in the limit of a perfect conducting plate, where $\mu=\infty$, the last term in (\ref{ETOTAL}) vanishes and the attenuation factor in the first terms equals to one.

On the right hand side of (\ref{ETOTAL}), the first term is proportional to the distance to the power $-1$ while the second term is proportional to the distance to the power $-3$. The last term exhibits a dependence on the distance to the power $-2$ so, for larger distances, this crossing term is more relevant in comparison with the direct magnetic dipole $\times$ image-magnetic dipole interaction.

For a quantum particle with spin $1/2$, the magnetic dipole is a quantum operator proportional to the spin operator ${\bf m}=\gamma{\bf S}$, where $\gamma$ is a constant of proportionality. In this sense, the Hamiltonian operator of the system is
\begin{eqnarray}
\label{defH}
H&=&\left[-\frac{q^{2}}{16\pi\mid b\mid}+\frac{\gamma^{2}\left[{\bf S}^{2}+\left(S_{z}\right)^{2}\right]}{64\pi\mid b\mid^{3}}\right]\left(\frac{\mu^{2}}{4+\mu^{2}}\right)\cr\cr
&\ &-\frac{q\gamma S_{z}}{8\pi\mid  b\mid^{2}}\left(\frac{\mu}{4+\mu^{2}}\right) \ .
\end{eqnarray}
The constant $\gamma$ is proportional to the charge $q$ and both have the same signal.

A general spin-$1/2$ state is given by
\begin{equation}
\label{defpsi}
|\psi\rangle=\cos\left(\theta/2\right)e^{-i\varphi/2}|+\rangle+\sin\left(\theta/2\right)e^{i\varphi/2}|-\rangle\ .
\end{equation}
where $\theta$ and $\varphi$ can be interpreted as the polar and azimuthal angels, respectively, associated with the mean spin direction of the state taken the origin at the particle position.

The mean value of the Hamiltonian operator (\ref{defH}) in the general spin-$1/2$ state (\ref{defpsi}) is
\begin{eqnarray}
\label{energiaquantica}
\langle\psi|H|\psi\rangle&=&\left[-\frac{q^{2}}{16\pi\mid b\mid}+\frac{\gamma^{2}}{64\pi\mid b\mid^{3}}\right]\left(\frac{\mu^{2}}{4+\mu^{2}}\right)\cr\cr
&\ &-\frac{q\gamma\cos(\theta)}{16\pi\mid  b\mid^{2}}\left(\frac{\mu}{4+\mu^{2}}\right)\ .
\end{eqnarray}

The first term on the right hand side of (\ref{energiaquantica}) leads to a Coulombian force between the particle and the plate, as discussed in section (\ref{III}). There is a remarkable difference between the second term in (\ref{energiaquantica}) and its classical counterpart. In the case of a quantum particle with spin, the contribution to the energy coming from the second term does not depend on the mean orientation of the spin with respect to the plate. In the case of a classical particle with magnetic moment, this contribution depends on the angle between the magnetic moment of the particle and the normal vector do the plate, as one can see in (\ref{ETOTAL}), with the scalar product ${\bf{m}}\cdot{\hat{z}}$.

The third term on the right hand side of (\ref{energiaquantica}) is the most interesting one. It is due to the magnetoelectric properties of the plate and is a $\theta$-dependent contribution, the mean angle between the spin of the state (\ref{defpsi}) and the normal to the plate. From the third term in (\ref{energiaquantica}) we have a mean contribution to the force between the plate and particle which falls with the distance between them slower in comparison with the second term in (\ref{energiaquantica}). Besides, this force also depends on the $\theta$ angle. 

In addition to a force, from the third term in (\ref{energiaquantica}), we also obtain a mean torque acting on the quantum particle with respect to the angle $\theta$. This torque is always non-positive, attains its minimum value for $\theta=\pi/2$ and vanishes when $\theta=0,\pi$.

\section{Conclusions}
\label{conclusoes}

The Maxwell electrodynamics in the vicinity of a semi-transparent mirror with magnetoelectric properties have been investigated. We have considered a mirror described  by a $\delta$-potential with a Chern-Simons-like coupling. We have found exactly the modification undergone by the photon propagator due to the presence of the mirror.  

We have computed the interaction energy between a stationary point-like charge and the mirror and we have shown that the role played by the coupling constant $\mu$ is just to attenuate the image charge by an uniform multiplicative factor. We have also shown that the interaction via image method is recovered in the limit case of a perfect mirror ($\mu\to\infty$).

We have shown that a stationary point-like charge produces an electric field and a magnetic field due to the presence of the mirror. The electric field is the one induced by the charge itself and the field associated to an image charge attenuated by a multiplicative factor. The magnetic field is the one associated with a magnetic monopole. The interesting magnetoelectric properties simulated by the model is an indication that it can be used to describe magnetoelectric surfaces.

We have investigated some physical phenomena which emerge from a setup composed by the plate and a stationary charged quantum particle with spin $1/2$. We have shown that, in addition a Coulombian and an isotropic dipole-dipole-like terms, the energy exhibits a magnetoelectric contribution with the features of a charge-magnetic dipole interaction. This last contribution leads also to a torque acting on the particle with spin. 

We hope that the present work could pave the way to use analytical methods to describe magnetoelectric material media. 
\newpage

{\bf Acknowledgements:}\\
For financial support,  L.H.C. Borges and H.L. Oliveira thank to CAPES (Brazilian agency) and F.A. Barone thanks to CNPq under the grants 311514/2015-4 and 313978/2018-2.

\appendix

\section{Stationary magnetic dipole source}

In this appendix we obtain the field source that describes a magnetic dipole ${\bf m}$. First we consider the vector potential related to the magnetic dipole placed at the origin 
\begin{equation}
\label{Adipolo}
{\bf A}({\bf x})=\frac{1}{4\pi}\frac{{\bf m}\times{\bf x}}{|{\bf x}|^{3}}\ .
\end{equation}

In a covariant form, the gauge field (\ref{Adipolo}) reads
\begin{equation}
\label{Amudipolo}
A^{\mu}_{MD}(x)=\frac{\epsilon^{0\mu j k}}{4\pi}\frac{m_{j}x_{k}}{|{\bf x}|^{3}}\ .
\end{equation}

As argued in reference \cite{Medeiros}, the Fourier transform of a gauge field, ${\tilde A}^{\mu}(p)$ is related to the Fourier transform of its corresponding field source, ${\tilde J}^{\mu}(p)$,
\begin{equation}
{\tilde A}^{\mu}(p)=-\frac{1}{p^{2}}{\tilde J}^{\mu}(p)\ .
\end{equation}

The Fourier transform of the field (\ref{Adipolo}) is obtained with the aid of (\ref{Amudipolo}), as follows
\begin{eqnarray}
\label{Amudipolotransformado}
&{\tilde A}^{\mu}_{MD}(p)&=\int d^{4}x A^{\mu}(x)e^{ipx}\cr
&=&\frac{\epsilon^{0\mu j k}}{4\pi}\left[\int dx^{0}e^{ip^{0}x^{0}}\right]\int d^{3}{\bf x}\frac{m_{j}x_{k}}{|{\bf x}|^{3}}e^{-i{\bf p}\cdot{\bf x}}\cr
&=&\frac{i\epsilon^{0\mu j k}}{2}\delta(p^{0})m_{j}\lim_{M\to\infty}\frac{\partial}{\partial p_{k}}\int d^{3}{\bf x}\frac{e^{-i{\bf p}\cdot{\bf x}}}{({\bf x}^{2}+M^{2})^{3/2}}\cr
&=&2\pi i\epsilon^{0\mu j k}\delta(p^{0})m_{j}\lim_{M\to\infty}\frac{\partial}{\partial p_{k}}K_{0}(|{\bf p}|M)\cr
&=&-2\pi i\epsilon^{0\mu j k}\delta(p^{0})m_{j}\frac{p_{k}}{|{\bf p}^{2}|}\ ,
\end{eqnarray}
where in the third line we introduced a mass parameter in order to regularize the integral.

With the aid of (\ref{Amudipolotransformado}) and (\ref{Amudipolo}) we have
\begin{equation}
{\tilde J}^{\mu}_{MD}(p)=-2\pi i\delta(p^{0})\epsilon^{0\mu j k}m_{j}p_{k}\ ,
\end{equation}
through which we can obtain the source, as follows
\begin{eqnarray}
\label{JDMintermediario}
J^{\mu}_{(MD)}(x)&=&\int\frac{d^{4}p}{(2\pi)^{4}}{\tilde J}^{\mu}(p)e^{-ipx}\cr\cr
&=&\epsilon^{0\mu j k}m_{j}\partial_{k}\int\frac{d^{3}{\bf p}}{(2\pi)^{3}}e^{i{\bf p}{\bf x}}\cr\cr
&=&\epsilon^{0\mu j k}m_{j}\partial_{k}\delta^{3}({\bf x})\ .
\end{eqnarray}

With a simple spatial translation, we can obtain the second term in (\ref{defJCMD}) from (\ref{JDMintermediario}) 

In vector form Eq. (\ref{JDMintermediario}) reads
\begin{equation}
{\bf J}({\bf x})=-{\bf m}\times\left[\nabla\delta^{3}({\bf x})\right]\ .
\end{equation}

\end{document}